\documentclass[12pt]{article}

\usepackage{amsmath,amsfonts}
\usepackage{amssymb}
\usepackage{epsfig,cite}
\usepackage{stmaryrd}
\usepackage[usenames,dvips]{color}
\usepackage{fullpage}
\usepackage{verbatim}

\makeatletter
\@addtoreset{equation}{section}
\makeatother

\newcommand{\be}{\begin{equation}}
\newcommand{\ee}{\end{equation}}
\newcommand{\bea}{\begin{eqnarray}}
\newcommand{\eea}{\end{eqnarray}}

\newcommand{\tr}{\operatorname{tr}}

\begin{document}

\thispagestyle{empty}

\begin{center}
\hfill BI-TP 2012/31\\
\hfill UAB-FT-714
\begin{center}

\vspace{.5cm}

{\Large\sc  Large diphoton Higgs rates from\\ supersymmetric triplets }

\end{center}

\vspace{1.cm}

\textbf{ Antonio Delgado$^{\,a}$, Germano Nardini$^{\,b}$
and Mariano Quir\'os$^{\,c}$}\\

\vspace{1.cm}
${}^a\!\!$ {\em {Department of Physics, University of Notre Dame\\Notre Dame, IN 46556, USA}}

\vspace{.1cm}
${}^b\!\!$ {\em {Fakult\"at f\"ur Physik, Universit\"at Bielefeld,
    D-33615 Bielefeld, Germany}}

\vspace{.1cm}

${}^c\!\!$ {\em {Instituci\'o Catalana de Recerca i Estudis  
Avan\c{c}ats (ICREA) and\\ Institut de F\'isica d'Altes Energies, Universitat Aut{\`o}noma de Barcelona\\
08193 Bellaterra, Barcelona, Spain}}

\end{center}

\vspace{0.8cm}

\centerline{\bf Abstract}
\vspace{2 mm}
\begin{quote}\small
  Recent results on Higgs searches at the LHC point towards the
  existence of a Higgs boson with mass of about $126$\,GeV whose
  diphoton decay rate tends to be larger than in the Standard
  Model. These results are in tension with natural MSSM scenarios:
  such a Higgs mass requires heavy (third-generation) squarks which
  reintroduce some amount of fine-tuning
  and in general the Higgs diphoton
  decay rate tends to follow the Standard Model result.
    In this paper we prove that these problems can be alleviated
  by introducing an extra supersymmetric triplet coupled to the Higgs
  in the superpotential. This superfield generates a sizeable
  tree-level correction to the Higgs mass so that the third generation
  is no longer required to be heavy, and its charged component enhances the
  diphoton Higgs decay rates by as much as $50\%$ with respect to the
  Standard Model values. We also show that such a scenario is compatible
  with present electroweak precision observables.
 
\end{quote}

\vfill

 \newpage



\textit{1. \underline{Introduction} } 
The ATLAS and CMS collaborations at CERN have recently reported~\cite{Gianotti:gia12, Incandela:inc12,
   ATLAS:2012ad,Chatrchyan:2012tw,Atlas,CMS}
   excesses in several channels compatible with a Higgs with mass $m_h\simeq 126$ GeV.
   This value of the Higgs mass puts a
 strong tension on the minimal supersymmetric extension of the
 Standard Model (MSSM) as very heavy third generation squarks and
 large stop mixing are required in order to reproduce it~\cite{Carena:2011aa}.
  Such mass values in the stop sector are in
 conflict with the MSSM as a natural solution to the hierarchy problem
 and create a \textit{little hierarchy} problem. 
 
 In fact in the MSSM
 the couplings of the Higgs sector are a prediction of the model so
 that in the decoupling limit the Standard Model (SM)-like Higgs mass turns out to
 be~\cite{Carena:1995bx}
\begin{eqnarray}
m^2_h&=& m^2_Z \cos^2 2\beta \left(1-\frac{3}{8 \pi^2}\frac{m_t^2}{v^2}\,t\right)\nonumber\\
&+&\frac{3}{4\pi^2}\frac{m_t^4}{v^2}\left[\frac{1}{2}X_t+t+\frac{1}{16\pi^2}\left(\frac{3}{2}\frac{m^2_t}{v^2}-32\pi\alpha_3\right)\left(X_tt+t^2\right)\right]~,
\label{Higgsmass}
\end{eqnarray}
with
\begin{eqnarray}
 X_t&=&\frac{2 \tilde{A}^2_t}{m_Q^2}\left(1-\frac{\tilde{A}^2_t}{12 m_Q^2}\right)~,\\
t&=&\log\left(\frac{m_Q^2}{m_t^2}\right)~,\qquad  \tilde{A}_t=A_t-\mu/\tan\beta~, 
\end{eqnarray}
where $m_t$ and $m_Z$ are the top and $Z$ masses, $\alpha_3$ is the
QCD coupling, $m_Q$ is the (common) supersymmetry breaking mass of the
third generation squarks~\footnote{For simplicity we will consider in
  this paper degenerate supersymmetry breaking masses $m_Q$ for up and
  down-type third generation squarks. Notice that perturbative
  problems can spoil the approximations used in Eq.~\eqref{Higgsmass}
  when $m_Q$ is very large in which case resumming logarithms is required.} 
  and $\mu$ is the holomorphic Higgsino
mass. Moreover, we use the notation $v=\sqrt{v_1^2+v_2^2}=174\,$GeV
and $\tan\beta=v_2/v_1$ where $v_1$ ($v_2$) is the vacuum expectation
value (VEV) of the Higgs $H_1$ ($H_2$) coupled to down (up) quarks. As
it was pointed out in Ref.~\cite{Carena:1995bx}, Eq.~(\ref{Higgsmass})
for values of $m_Q\lesssim 1.5$~TeV provides a good approximation
(within 2 GeV error) to more sophisticated numerical results.
Actually one easily obtains from Eq.~\eqref{Higgsmass} that quite
heavy third generation squarks with $m_Q=\mathcal O$(1\,TeV), large
$\tan\beta$ and a sizeable mixing $X_t$ are
needed~\cite{Carena:2011aa} in order to reproduce a SM-like Higgs mass
of about 126\,GeV. These heavy squarks however induce large radiative
contributions to the electroweak breaking mechanism and then tend to
rise the electroweak scale far away from the observed one (say the $Z$
mass) unless a fine-tuning of around one per mille is done.

In short, by taking naturalness as a guiding criterion for physics
beyond the SM, there is a tension in the MSSM between
the actual values of the Higgs and $Z$ boson masses.  To alleviate
this tension a simple supersymmetric possibility (without
enlarging the SM gauge group) is to extend the MSSM with some extra
multiplets which trigger extra tree level contributions to the Higgs
quartic coupling, in such a way that one could reproduce the
experimental value of the Higgs mass without the need of heavy
third-generation squarks. Several proposals have been made in the
literature~\cite{Espinosa:1991gr}~\footnote{In this paper we focus on
  low-energy extensions of the MSSM. For non-minimal ultraviolet
  completions with the extra content at the multi-TeV scale, see for
  instance \cite{highscale}.}. As the extra multiplet has to be
coupled to the Higgs sector in the superpotential by renormalizable
couplings, the number of possible extra multiplets is reduced: either
$SU(2)_L$ singlets or triplets with hypercharge $Y=0$ or $\pm 1$ can
play this role.

On the other hand, a further source of tension comes from the LHC
measurements of the Higgs decay rates. From the experimental results
on the Higgs decay into $ZZ$ and $WW$ channels one may infer that in
this sector no dominant contributions beyond the SM ones do exist in Nature,
as well as in the Higgs production through either gluon or weak vector
boson fusion. However new physics beyond the SM
will arise from the diphoton Higgs decay rate 
if LHC keeps on showing a significant excess with respect to the SM prediction
when more LHC data will be
collected. Assuming SM-like Higgs production, the
ratio between the diphoton rate observed at LHC and the one expected
in the SM is $R_{\gamma\gamma}=1.8\pm 0.5$ for ATLAS ($m_h=126$ GeV)~\cite{Atlas} and
$R_{\gamma\gamma}=1.6\pm 0.4$ for CMS ($m_h=125$ GeV)~\cite{CMS}. The central value of the
combination of these measurements, which roughly corresponds to an
enhancement of $\sim 1.7$ with respect to the SM prediction (with an error around $\pm 0.3$) is hard to
reproduce in the MSSM~\cite{Carena:2011aa} and, if this central value
were confirmed with more statistics, it would represent a strong
tension between the MSSM and LHC data.
A supersymmetric extension of the SM solving this problem should then
extend the MSSM by some electrically charged (extra) states coupled to
the Higgs that should contribute to $R_{\gamma\gamma}$ at one-loop. 
Notice that these extra states, possibly charged
under both $SU(2)_L$ and $U(1)_Y$, would generate a subleading (one-loop)
correction to the SM (tree-level) weak vector fusion
production. Moreover they should be colorless in order to not modify
at leading order the gluon-fusion Higgs production arising at one-loop
in the SM.

In the present paper we analyze minimal MSSM extensions where extra
states can relax the little hierarchy problem in the presence of a 126
GeV Higgs mass as well as reproduce the diphoton excess in the Higgs
production rate. 

\textit{2. \underline{The model} } As the singlet is electrically neutral only triplets,
which contain charged states and can thus contribute to the
$h\to\gamma\gamma$ decay width, are natural candidates to the MSSM
extension. In particular we consider the effect of a supersymmetric
$Y=0$ triplet~\footnote{Considering supersymmetric triplets with $Y=\pm 1$ should lead to similar results as those founds in the present paper.}
\be
\Sigma=\left(
\begin{array}{cc}
  \xi^0/\sqrt{2} & -\xi_2^+\\
  \xi_1^-&  -\xi^0/\sqrt{2}
\end{array}
\right)
\label{Sigma}
\ee
on both issues: the Higgs mass generation and the diphoton rate. 

The most general renormalizable coupling of the triplet $\Sigma$ to
the Higgs sector is provided by the superpotential
\begin{equation}
\label{super}
\Delta W=\lambda H_1 \cdot \Sigma H_2+\frac{1}{2} \mu_\Sigma \tr\Sigma^2
\end{equation}
that gets added to the MSSM one. Notice that $\tr\Sigma^3\equiv 0$ due
to its own structure. The new interaction modifies the Higgs potential
and, in the decoupling limit, the tree-level mass of the SM-like Higgs
is given by
%
\begin{equation}
\label{tree-level}
m^2_{h,{\rm tree}}=m_Z^2\cos^2 2\beta+\frac{\lambda^2}{2} v^2 \sin^2 2\beta~.
\end{equation}
We see that for moderate values of $\lambda$ the tree-level mass can
be lifted so that no large contributions from loop corrections are
required to reproduce $m_h\simeq 126\,$GeV. In particular, stops can be
light, thus reducing the fine tuning for the electroweak scale. Notice
also that the new contribution is relevant mostly when
$\tan\beta\simeq 1$ while it vanishes when $\tan\beta\to\infty$. In
this way it will be possible to cope with the LHC Higgs mass for
moderate values of $\tan\beta$ without large stop mixing, contrarily
to what happens in the MSSM.

One strong constraint on models with triplets comes from the
electroweak precision tests (EWPT), in particular from the
$\rho$-parameter constraint~\cite{Nakamura:2010zzi}. If the scalar component of the
triplet acquires a VEV $\langle \xi^0\rangle$ it will give a
tree-level contribution to the $\rho$ parameter as $\rho=1+2\langle
\xi^0\rangle^2/v^2$~\cite{Nakamura:2010zzi} which will easily be in
conflict with experimental data. On the other hand a triplet VEV will
always exist in a theory with a superpotential given by
Eq.~(\ref{super}) after electroweak breaking. Moreover once supersymmetry is broken, one expects
the soft-breaking term $\lambda A_\lambda H_1\cdot \Sigma H_2$ to
appear in the scalar potential, depending on the particular mechanism of supersymmetry breaking, where now all symbols denote just the
scalar components of the chiral superfields. So once the Higgs gets a VEV
it will generate a tadpole for the neutral component $\xi^0$ of
$\Sigma$ and the previous terms will induce the VEV
\be
\langle \xi^0\rangle\simeq\sqrt{2} \lambda \left[\mu+\frac{1}{2}\left(\mu_\Sigma+\frac{A_\lambda}{2}\right)\sin 2\beta\right] \frac{v^2}{m_\Sigma^2+\mu_\Sigma^2+\lambda^2 v^2/2}~,
\ee
where $m_\Sigma$ is the supersymmetry breaking mass for the triplet
and $A_\lambda$ is the trilinear supersymmetry breaking parameter
associated to the superpotential (\ref{super}). The present bound on the
$\rho$ parameter,
$\rho=1.0004^{+0.0003}_{-0.0004}$~\cite{Nakamura:2010zzi}, imposes the
constraint $\langle \xi^0\rangle\lesssim 4$ GeV at 95\% CL. In order
to be consistent with the experimental value of the $\rho$ parameter we
are going to suppose that there is a hierarchy between the trilinear coupling $A_\lambda$ and supersymmetric masses in the superpotential (\ref{super}), and
the soft mass for the $\Sigma$-scalar~\footnote{Notice that the required hierarchy $A_\lambda\ll m_\Sigma$ naturally arises if the supersymmetry breaking mechanism is driven by gauge interactions where both $m_\Sigma^2$ and $A_\lambda$ are obtained at two-loop.}, i.e.~$A_\lambda,\,\mu,\,\mu_\Sigma\ll m_\Sigma$. Moreover to cope with the
experimental value of the $\rho$ parameter we will consider that the scalar component is
in the TeV range and thus its presence is negligible in this
discussion. We will then hereafter neglect $\langle\xi^0\rangle$ and
put it to zero.

In the fermion sector $\tilde\xi^0$ mixes with the MSSM neutralinos
$(\widetilde W^3,\widetilde W^0,\widetilde H_1^0,\widetilde H_2^0)$
while $\widetilde \xi_1^-$ and $\widetilde \xi_2^+$ mix with the MSSM
charginos $(\widetilde W^{-},\widetilde W^+,\widetilde
H_1^-,\widetilde H_2^+)$. As the relevant states for the ratio
$R_{\gamma\gamma}$ are the charged ones we will concentrate on
charginos. Their mass matrix is given by
\be
\left(
\widetilde W^-,~\widetilde H_1^-,~\widetilde \xi_1^-
\right)
\mathcal M_{ch}
\left(
\begin{array}{c}\widetilde W^+ \\ \widetilde H_2^+\\\widetilde \xi_2 ^+
\end{array}
 \right),\quad \mathcal M_{ch}=\left(
\begin{array}{ccc} M_2& gv\sin\beta& 0\\ gv\cos\beta&\mu&\lambda v\sin\beta\\0&\lambda v\cos\beta& \mu_\Sigma\end{array}
\right)~,
\label{charginos}
\ee
where we have put for simplicity $\langle\xi^0\rangle=0$ while for
definiteness the parameters $M_2,\mu$ and $\mu_\Sigma$ will be assumed to be
positive in the rest of the paper. In the limit where $M_2\gg m_W$,
the three charged eigenstates $(\widetilde \chi_1^+,\widetilde
\chi_2^+,\widetilde \chi_3^+)$ have respectively masses
$(m_{-},m_+,M_2)$ where
\begin{eqnarray}
m_{\pm}&=&\mu_+\pm\sqrt{v_1 v_2\lambda^2+\mu_-^2}~,\nonumber\\
\mu_\pm&=&\frac{\mu\pm \mu_\Sigma}{2}~.
\label{masses}
\end{eqnarray}
Therefore the parameter $m\equiv m_-$ is a good estimate for the mass
of the lightest chargino $m_{\widetilde\chi_1^\pm}$, which we will
impose to be heavier than 94\,GeV~\cite{Nakamura:2010zzi}, and the
whole chargino sector can be expressed as function of
$m,\lambda,\tan\beta,\mu_-$ and $M_2$. In all cases we will choose the value of the mass parameter
$m$ such that, in the corresponding region of the parameter space, the experimental bounds on the 
chargino masses are fulfilled. 


\textit{3. \underline{The ratio $h\to\gamma\gamma$} } In the limit where $m_h^2\ll 4m_{\widetilde \chi_i^{+}}^2$ and by
taking into account the main contributions due to charginos, $W$ boson
and top quark $t$, the diphoton Higgs decay rate with respect to the SM value
$R_{\gamma\gamma}$ turns out to be~\cite{Ellis:1975ap,Shifman:1979eb,Carena:2012xa}
\be
R_{\gamma\gamma}=\left| 
1+\frac{ {\displaystyle \frac{4}{3}\frac{\partial}{\partial \log v} } \log\det \mathcal M_{ch}(v)
}{A_1(\tau_W)+{\displaystyle \frac{4}{3} } A_{1/2}(\tau_t)}
\right|^2~,
\label{R}
\ee
where $\tau_i=m_h^2/4 m_i^2$. The functions $A_{1}(\tau)$ and
$A_{1/2}(\tau)$ are given in Ref.~\cite{Djouadi:2005gj} and their numerical values
for the $W$ and top fields are $A_1(\tau_W)\simeq -8.3$ and
$A_{1/2}(\tau_t)\simeq 1.4$ so that the denominator in Eq.~(\ref{R})
is negative. The
numerator in Eq.~(\ref{R}) is given by
\be
\frac{\partial}{\partial \log v}  \log\det \mathcal M_{ch}(v)=-\frac{\sin 2\beta v^2(\lambda^2 M_2+g^2\mu_\Sigma)}
{M_2\mu\mu_\Sigma-\frac{1}{2}\sin 2\beta v^2(\lambda^2 M_2+g^2\mu_\Sigma)}~,
\label{num}
\ee
and its sign depends on the specific values that one can choose for
the parameters~\footnote{Notice that forbidding massless charginos
  imply finiteness of the numerator in Eq.~(\ref{R}).}. In the present
paper we are interested in cases where the r.h.s.~of Eq.~\eqref{num}
is positive, which leads to an enhancement of the diphoton Higgs decay
rate as suggested by present LHC data.  This will be studied by
imposing $m_h=126$\,GeV.  In particular as we do not need very heavy
third-generation squarks to achieve such a Higgs mass, in the
following illustrative examples we will fix $m_Q\simeq 700$, $X_t=0$ (i.e.~$m_{\tilde t_1}=m_{\tilde t_2}=700$ GeV) and $X_t=4$  (i.e.~$m_{\tilde t_1}=545$ GeV and $m_{\tilde t_2}=828$ GeV).

\begin{figure}[htb]
\begin{center}
\includegraphics[width=0.49\textwidth]{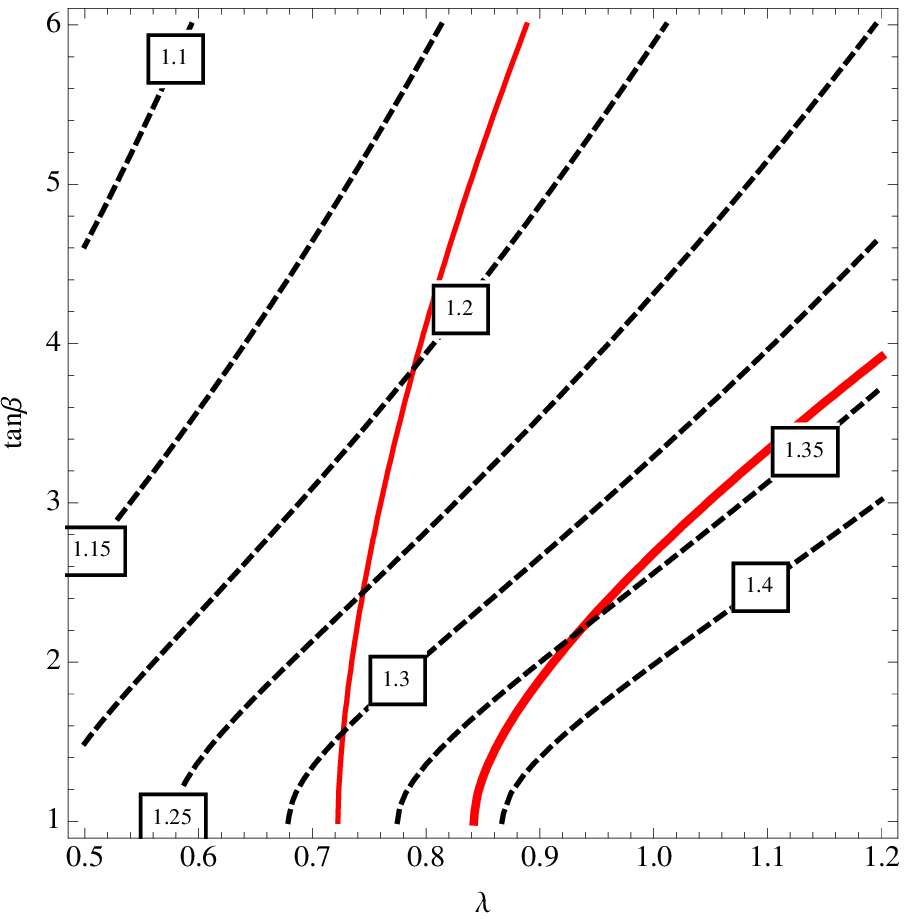}
\includegraphics[width=0.49\textwidth]{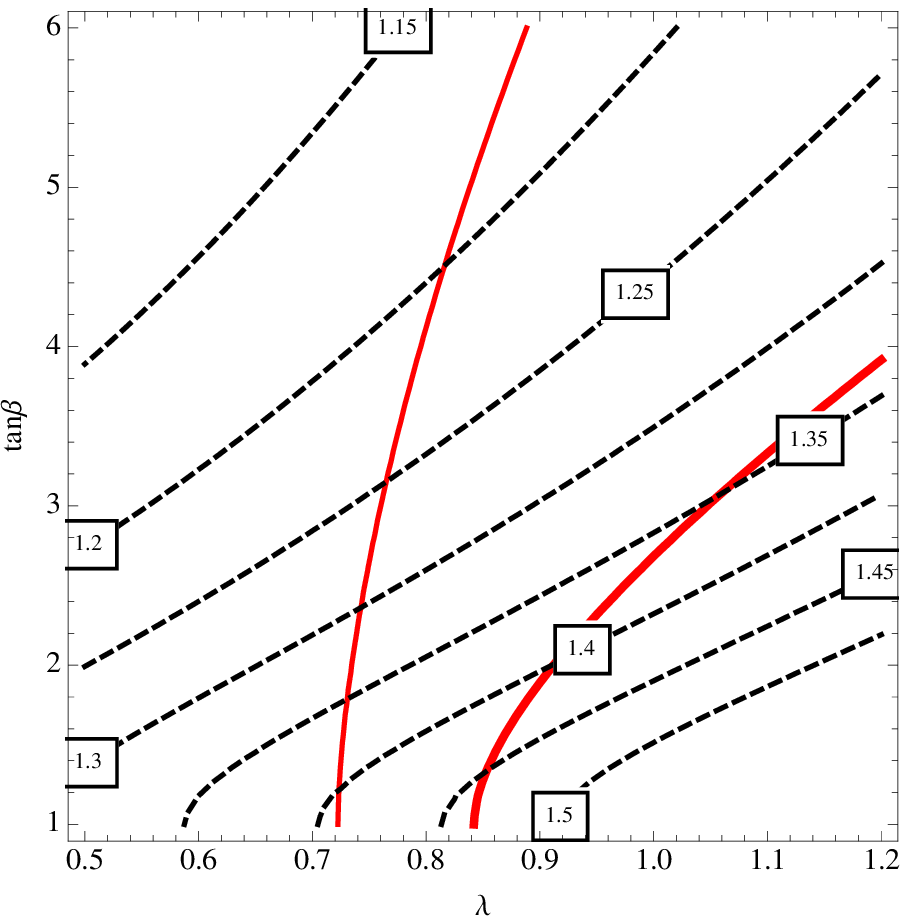}
\end{center}
\caption{\it Left panel: Contour plots of $R_{\gamma\gamma}$ (dashed lines) for $M_2=750$ GeV,  $m=100$ GeV and  $\mu_-=0$. Contour plots of $m_h=126$ GeV for $X_t=0$ [thick solid (red) line] and for $X_t=4$ [thin solid (red) line]. Right panel: The same as in the left panel but for $M_2=250$ GeV and $m=117$ GeV.}
\label{ratio}
\end{figure}

In Fig.~\ref{ratio} we present the contour lines of
$m_h=126$ GeV for $X_t=0$ [thick solid (red) curves] and $X_t=4$ [thin
solid (red) curves] and the contour lines of $R_{\gamma\gamma}$ (dashed
black curves) in the plane ($\lambda,\tan\beta$) where we fix
$\mu=\mu_\Sigma$. In the left panel, where we take $M_2=750$\,GeV and
$m=100\,$GeV, the chargino masses vary as $m_{\widetilde\chi_1^+}\in[95,110]$\,GeV,
$m_{\widetilde\chi_2^+}\in[217,302]$\,GeV and
$m_{\widetilde\chi_3^+}\in[754,762]$\,GeV.
As we can see the considered model
cannot reach $R_{\gamma\gamma}\simeq1.4$ for mixing $X_t=4$ but it
can for $X_t=0$, $\lambda\simeq 0.85$ and $\tan\beta\simeq 1$. Of
course, larger values of $R_{\gamma\gamma}$ can be achieved by
decreasing $M_2$ a possibility shown
in the right panel of Fig.~\ref{ratio} where $M_2=250$\,GeV and
$m=117$\,GeV are fixed. In such a case, the model can roughly reach the value $R_{\gamma\gamma}\simeq 1.5$ for
$X_t=0$ in correspondence with $\tan\beta\simeq 1$
and $\lambda\simeq 0.85$. In the region $(\lambda,\tan\beta)$ shown in the right panel of Fig.~\ref{ratio} the masses of the
charginos vary as $m_{\widetilde\chi_1^+}\in[95,110]$\,GeV,
$m_{\widetilde\chi_2^+}\in[215,240]$\,GeV and
$m_{\widetilde\chi_3^+}\in[280,360]$\,GeV.

\begin{figure}[htb]
\begin{center}
\includegraphics[width=0.49\textwidth]{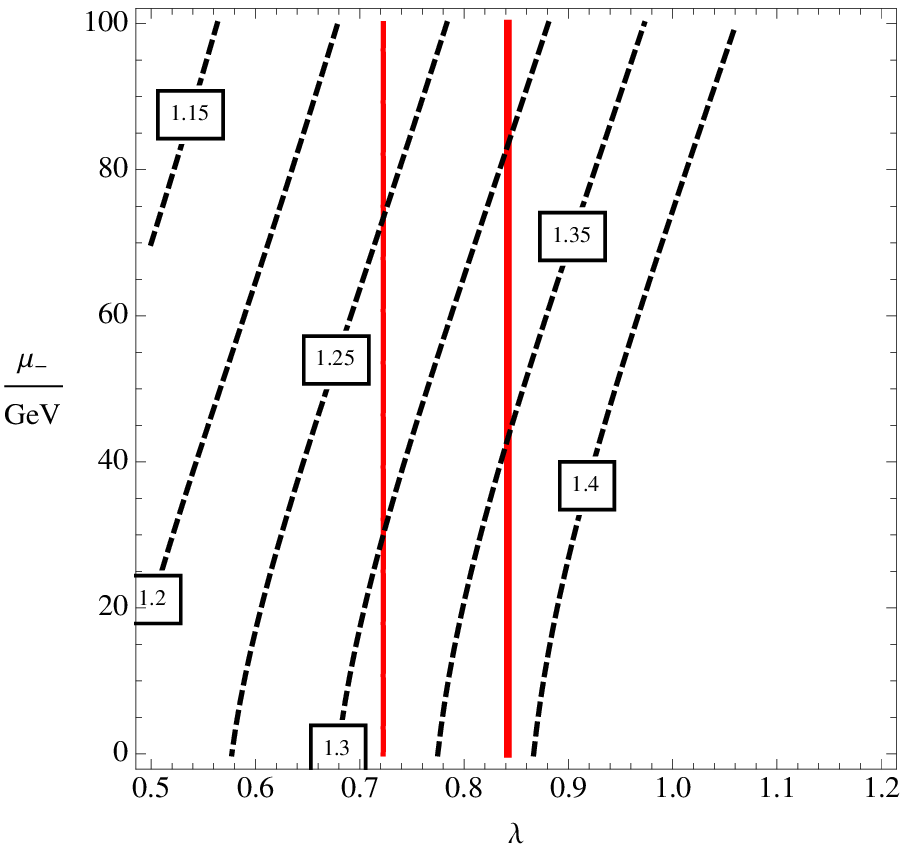}
\includegraphics[width=0.49\textwidth]{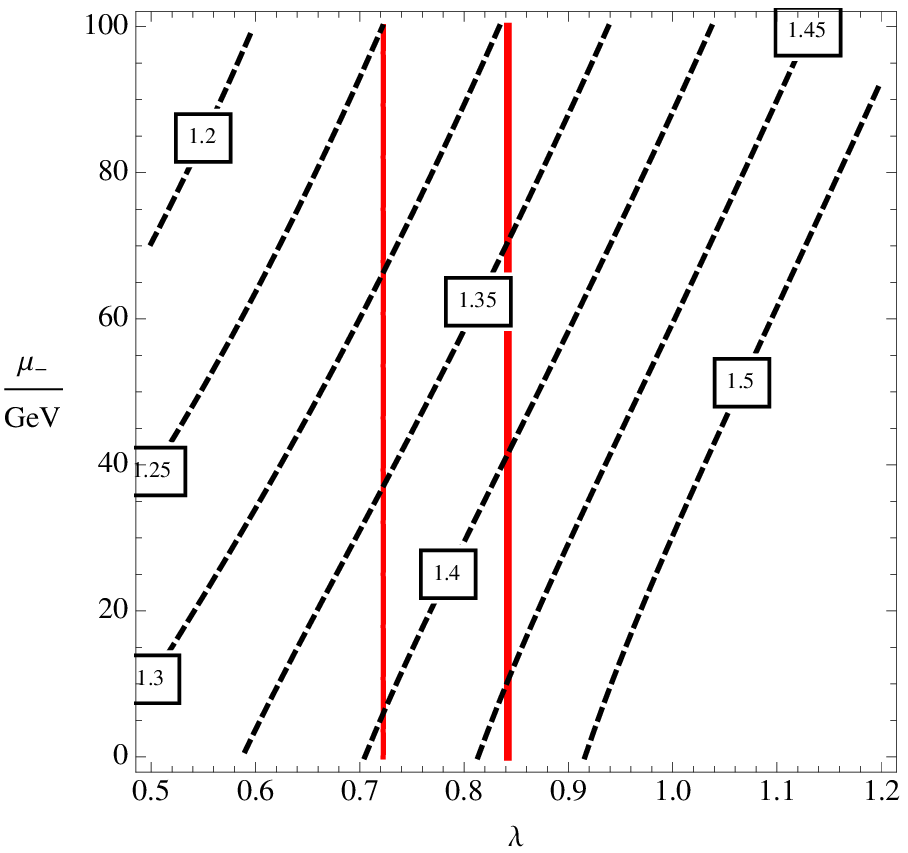}
\end{center}
\caption{\it Left panel: Contour plots of $R_{\gamma\gamma}$ (dashed
  lines) for $M_2=750$ GeV, $m=100$ GeV and $\tan\beta=1$ in the plane
  $(\lambda,\mu_-/\textrm{GeV})$ and $m_h=126$ GeV for $X_t=0$ [thick
  solid (red) line] and $X_t=4$ [thin solid (red) line]. Right panel: Same as in the left panel but for
  $M_2=250$ GeV and $m=117$ GeV.}
\label{muminus}
\end{figure}
Up to now we have presented results on $R_{\gamma\gamma}$ for
$\mu_-=0$. The variation with $\mu_-$ for $\tan\beta=1$ is shown in
Fig.~\ref{muminus} where we vary $\mu_-$ for $M_2=750$ GeV, $m=100$
GeV (left panel) and $M_2=250$ GeV, $m=117$ GeV (right panel).  We can
see that increasing $\mu_-$ (starting from $\mu_-=0)$ does not enhance
$R_{\gamma\gamma}$ for a given Higgs mass curve (thin red line for
$X_t=4$ and thick red line for $X_t=0$). Instead, for $\mu_-<0$ one
might naively extrapolate from the figure that negative values of
$\mu_-$ lead to better results, but for the shown choices of
parameters such a possibility would actually yield too small lightest
chargino masses.

\begin{figure}[htb]
\begin{center}
\includegraphics[width=0.49\textwidth]{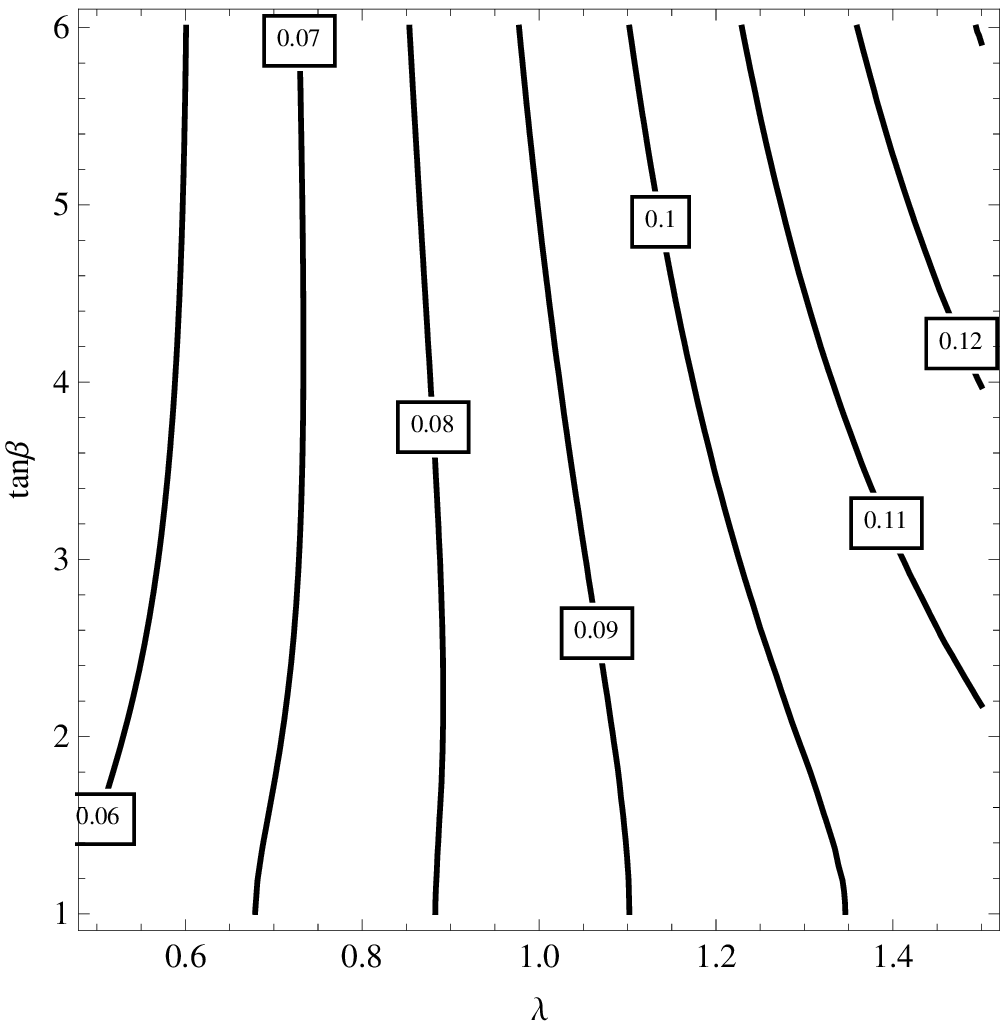}
\includegraphics[width=0.49\textwidth]{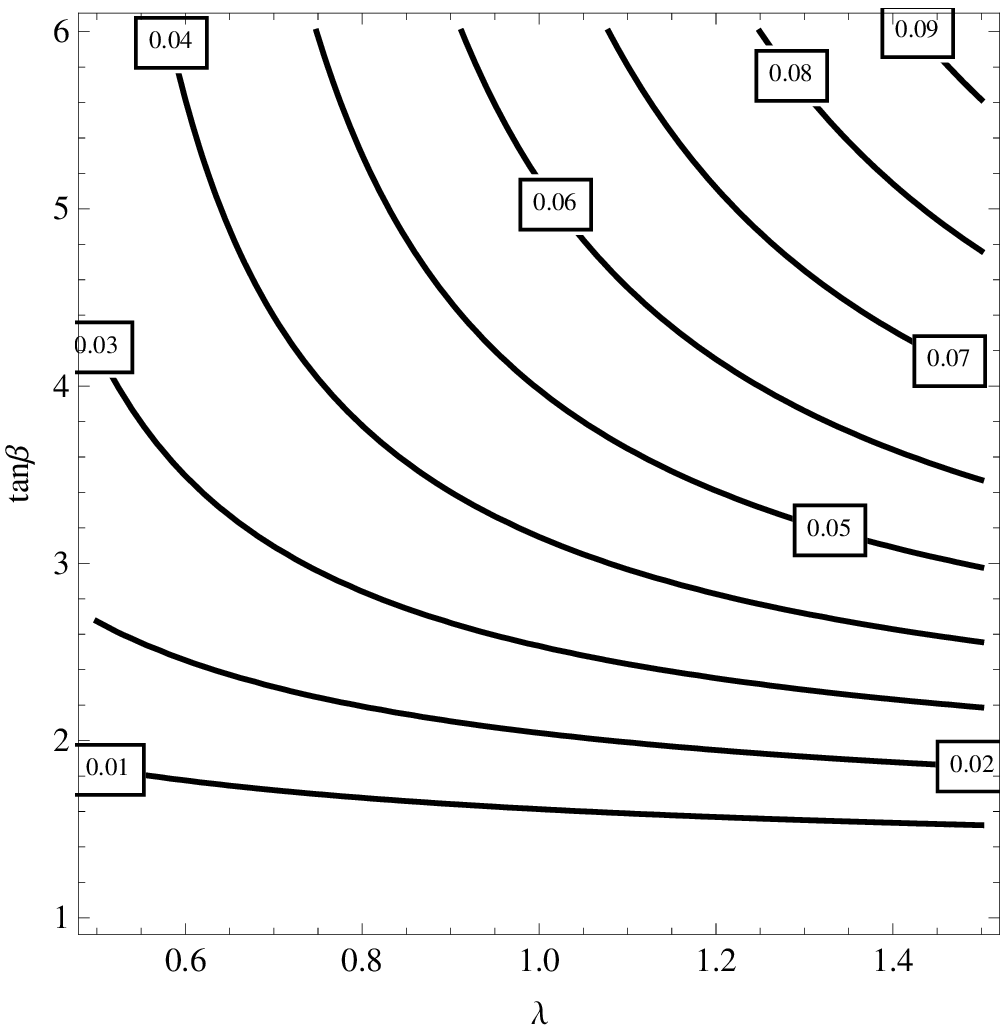}
\end{center}
\caption{\it Contour plots of the S (left panel) and T (right panel) parameters in the plane $(\lambda,\tan\beta)$ for $m=117$ GeV, $M_2=250$ GeV and $\mu=\mu_\Sigma$.}
\label{fig-ST}
\end{figure}
\textit{4. \underline{Electroweak observables} } An important question is how the zero-hypercharge supersymmetric
triplet modifies the electroweak observables, especially the $T$ parameter~\cite{Peskin:1991sw}
which is particularly sensitive to the presence of the
triplet. Indeed, as already mentioned above, the triplet contributes
to the $T$ parameter at tree-level as the previously considered $\rho$-parameter and $T$ are related by $\rho-1=\alpha T$, where $\alpha$ is the electromagnetic constant at the $m_Z$ scale. As we already mentioned electroweak breaking produces at tree-level a
tadpole in its neutral component $\xi^0$ and the experimental
constraint on this contribution then requires
$\langle\xi^0\rangle\lesssim 4$ at 95\% CL. 

Moreover at one-loop the
supersymmetric triplet contributes to the electroweak observables
through its coupling to the Higgs sector and, for $\mu=\mu_\Sigma$,
the oblique $S$ and $T$ parameters~\cite{Peskin:1991sw} get modified compared to the MSSM as
\begin{eqnarray}
\alpha S&=&\frac{s_W^2\lambda^2}{10\pi^2}\,\frac{m_W^2}{\mu^2}\left[1+\frac{19}{24}\sin 2\beta\right]+\mathcal O(g^4)~,\nonumber\\
\alpha T&=& \frac{3\lambda^2}{128\pi^2}\,\frac{m_W^2}{\mu^2}\, \cos^2 2\beta+\mathcal O(g^4)~,
\label{ST}
\end{eqnarray}
where the $\mathcal O(g^4)$ correction is coming from the (MSSM) Wino-Higgsino mixing.
These corrections are based on an expansion at ${\mathcal
  O}(s_W^2)$~\cite{Marandella:2005wc}  and turn out to be small in the considered region as compared with the experimental values~\cite{Nakamura:2010zzi}
\be  
S=0.04\pm 0.09,\quad T=0.07\pm 0.08\qquad  \textrm{(88\% correlation)}~.
\label{expST}
\ee

We plot in Fig.~\ref{fig-ST} contour lines for the predicted values of $S$ (left panel) and $T$ (right panel)
for $m=117$\,GeV, $M_2=250$ GeV and $\mu=\mu_\Sigma$. We have included on top of the triplet contributions from Eq.~(\ref{ST}) the $\mathcal O(g^4)$ contribution coming from the MSSM gaugino-Higgsino mixing. The predicted values of the $S$ and $T$ parameters are in all cases in agreement  with 
the experimental values within 1$\sigma$. Moreover, in the region where
$m_h=126\,$GeV (for $m_Q=700\,$GeV) and the diphoton rate is maximally
enhanced (i.e. $\tan\beta\simeq 1$ and $\lambda\simeq 0.85$) we obtain
$S\simeq 0.08$ and $T\simeq 0$.
The technical reasons why triplet corrections to the electroweak
observables are generally small in the considered region are:
\begin{itemize}
\item
For $\tan\beta=1$ (and zero triplet VEV) the custodial symmetry is
unbroken by the triplet and thus $T=0$.
\item Apart from the loop factor there is the extra suppression
  $m_W^2/\mu^2$ and it turns out that this ratio is small. For
  instance with $m=117$ GeV, $\lambda=0.85$ and $\tan\beta=1$ we get
  $\mu=221$ GeV for $\tan\beta=1$ and $\mu=177$ GeV for $\tan\beta=6$.
\end{itemize}

\begin{figure}[htb]
\begin{center}
\vspace{.5cm}
\includegraphics[width=0.50\textwidth]{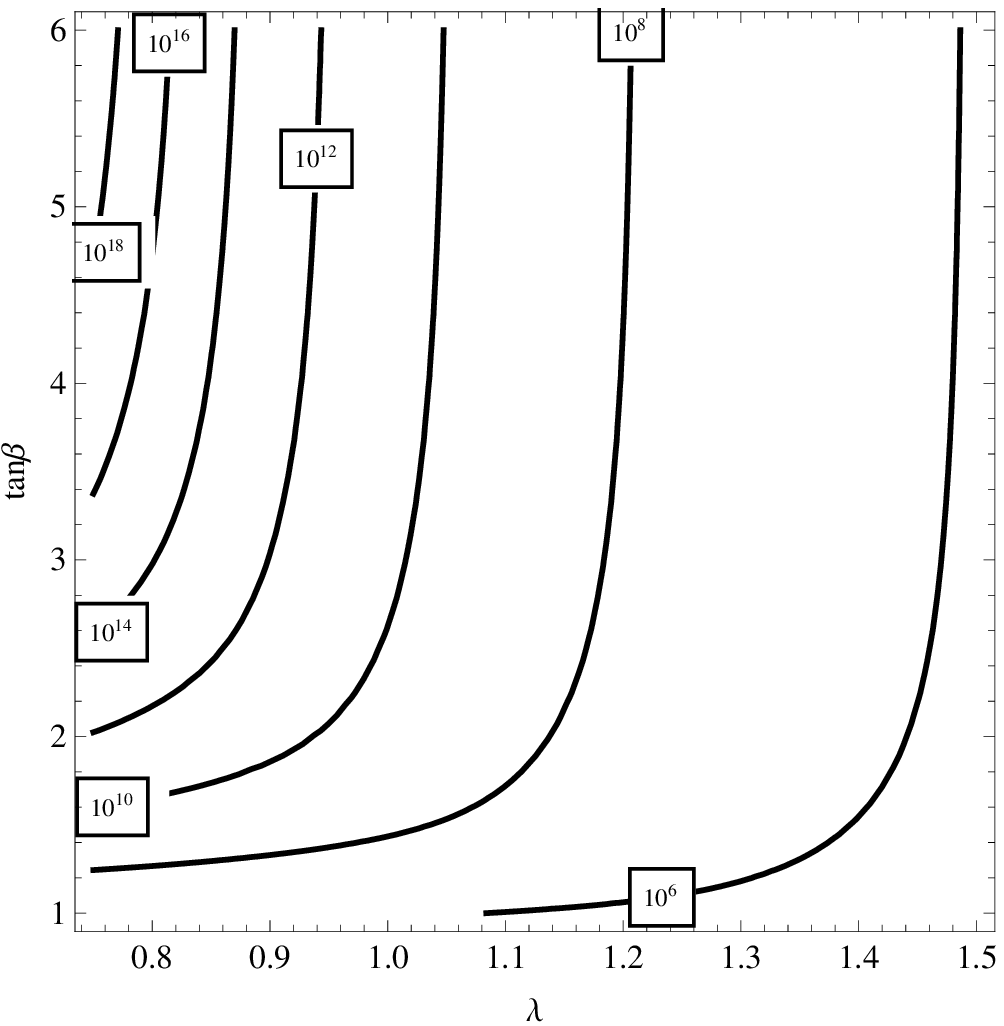}
\includegraphics[width=0.49\textwidth,height=6.5cm]{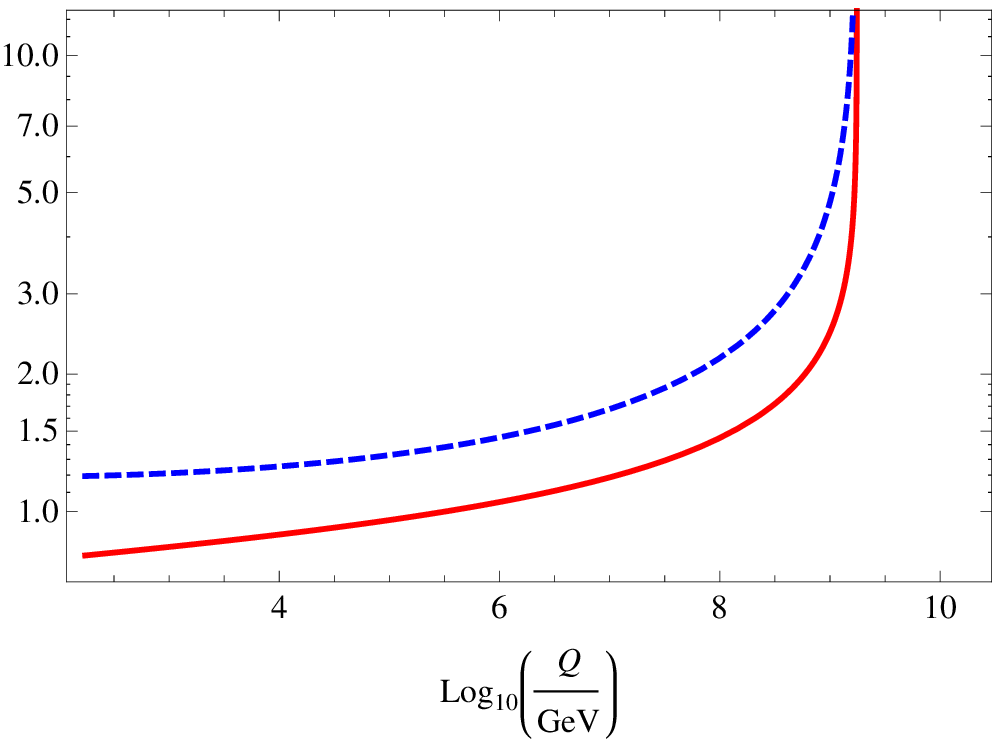}
\end{center}
\caption{\it Left panel: Contour lines for constant values of the
  cutoff $\Lambda$ (in GeV) in the plane $(\lambda,\tan\beta)$. Right
  panel: Plot of $\lambda(t)$ [solid (red) line] and $h_t(t)$ [dashed (blue) line]
  for $\tan\beta=1.5$ and $\lambda=0.8$.}
\label{Landau}
\end{figure}

\textit{5. \underline{Perturbativity} } A final issue which has to be considered is perturbativity of coupling
constants. In fact the evolution with the scale of the couplings
$\lambda$ and $h_t$ are given by the renormalization group equations
(RGE)~\cite{{Espinosa:1991gr}}
\begin{eqnarray}
8\pi^2\dot{\lambda}&=&\left(-\frac{7}{2}g^2-\frac{1}{2}g'^2+2\lambda^2+\frac{3}{2}h_t^2  \right)\lambda~,\nonumber\\
8\pi^2\dot{h}_t&=&\left(-\frac{3}{2}g^2-\frac{13}{18}g'^2-\frac{8}{3}g_3^2+\frac{3}{4}\lambda^2+3h_t^2  \right)h_t~,\label{rge}\\
16\pi^2\dot{g}&=&3g^2~,\quad 16\pi^2\dot{g}'=11g'^2~,\quad
16\pi^2 \dot{g}_3=-3g_3^3~,\nonumber
\end{eqnarray}
where the dots stand for $d/dt$ and $t=\log (\mathcal
Q/\textrm{GeV})$. We can see from the first equality in
Eq.~(\ref{rge}) that for large enough initial values of $\lambda\equiv
\lambda(m_t)$, the running coupling $\lambda(\mathcal Q)$ is driven to
larger values at high scales and eventually it reaches
non-perturbative values ($\lambda(\mathcal Q)\simeq 4\pi$) in the
ultraviolet (UV) at some scale $\mathcal Q\simeq \Lambda$, the cutoff
of the theory, near its Landau pole. This means that the theory
becomes non-perturbative, unless it is UV completed at some scale
smaller than $\Lambda$.  It is thus clear that the MSSM with an extra
zero-hypercharge triplet does not unify
perturbatively~\footnote{Indeed, even in the regime where $\lambda$ is
  small so that the theory remains perturbative until the Planck
  scale, unification cannot be achieved because the extra triplet
  modifies the MSSM beta functions of the gauge coupling in an
  incomplete way.}. However if the theory is still valid near its
Landau pole all couplings will feel the singularity through the higher
loop RGE and they can unify through the so-called non-perturbative
unification~\cite{Maiani:1977cg}~\footnote{See
  Ref.~\cite{Espinosa:1991gr} for examples of non-perturbative
  unification applied to the extensions of the MSSM with additional
  matter.}. With respect to conventional unification, non-perturbative
unification has the attractive feature that low energy couplings are
less sensitive to high energy physics. Fig.~\ref{Landau} (left panel)
shows the value of the cutoff $\Lambda$ (in GeV) in the plane
$(\lambda,\tan\beta)$ for the MSSM model completed with the
supersymmetric $Y=0$ triplet $\Sigma$.
We can see that when $\lambda=0.8$ the cutoff $\Lambda$ ranges from
$10^8$ GeV to $10^{16}$ GeV for $\tan\beta\simeq1$ to
$\tan\beta\simeq 6$, respectively. In particular, in the parameter region considered
in the right panels of Figs.~\ref{ratio} and
\ref{muminus}, for $R_{\gamma\gamma}\simeq 1.5$
 and $m_h=126\,$GeV it turns out that $\Lambda\simeq 2\times
10^9$ GeV. This can be also seen in the right panel of
Fig.~\ref{Landau} where the evolution of the couplings $\lambda(t)$
and $h_t(t)$ is presented.

\textit{6. \underline{Conclusion} } In conclusion if one interprets the excess discovered at LHC as a
Higgs boson with a mass about 126\,GeV, a little hierarchy problem
emerges in the MSSM. In addition a further tension between the MSSM and
experimental results would arise if the actual tendency of ATLAS and CMS data, on Higgs production and decay, 
towards deviations from the SM results only in the Higgs diphoton
decay channel, would be confirmed with better precision in the present (and forthcoming) LHC run.
In this paper we have proven that both problems can be naturally overcome by minimally
extending the MSSM by a colorless zero-hypercharge $SU(2)_L$-triplet
with a coupling to the Higgs superfields of order one, in fact $\lesssim h_t$, the top quark Yukawa coupling. In such a case,
even for small $\tan\beta$ and no mixing between the third-generation
squarks, it is possible to reach the experimental value of the Higgs mass $m_h=126\,$GeV and
at the same time $\sim$ 50\% enhancement in the Higgs diphoton decay
rate. In the considered parameter range the theory is consistent with electroweak precision tests and hints towards a
non-perturbative unification at the scale $\sim \,10^9$\,GeV. Finally the experimental signals
of this triplet could come from pair production of neutralinos (charginos) which then
decay mainly into Higgs plus missing energy.
 

\section*{\sc Acknowledgments}
  AD was partly supported by the National Science Foundation
under grants PHY-0905383-ARRA and PHY-1215979. MQ was supported by the Spanish
Consolider-Ingenio 2010 Programme CPAN (CSD2007-00042) and by
CICYT-FEDER-FPA2008-01430 and FPA2011-25948.

\end{document}